\documentclass[showpacs,amssymb,prc]{revtex4}
\usepackage[dvips]{graphicx}
\def\ds{\displaystyle}

\def\bmath#1{\mbox{\boldmath $#1$}}
\newcommand{\clebschgordan}[6]{ \langle \, #1 \, #2  \, ; \, #3 \, #4 \, | \,
                                                       #5 \, #6 \, \rangle }

\begin{document}

\title{Spin and pseudospin symmetries and the equivalent spectra  of
relativistic spin-1/2 and spin-0 particles}
\author{P. Alberto}
\affiliation{Physics Department and Center for Computational
Physics, University of Coimbra, P-3004-516 Coimbra, Portugal}
\author{A. S. de Castro} \affiliation{Departamento de
F\'{\i}sica e Qu\'{\i}mica, Universidade Estadual Paulista,
 12516-410 Guaratinguet\'a, SP, Brazil}
\author{M. Malheiro}
\affiliation{Departamento de F\'{\i}sica, Instituto Tecnol\'ogico de
Aeron\'autica, CTA, 12228-900, S\~ao Jos\'e dos Campos, SP, Brazil
\\ and Instituto de F\'{\i}sica, Universidade Federal Fluminense,
24210-340 Niter\'oi, Brazil}


\date{\today}


\begin{abstract}
\noindent We show that the conditions which originate the spin and
pseudospin symmetries in the Dirac equation are the same that
produce equivalent energy spectra of relativistic spin-1/2 and
spin-0 particles in the presence of vector and scalar potentials.
The conclusions do not depend on the particular shapes of the
potentials and can be important in different fields of physics. When
both scalar and vector potentials are spherical, these conditions
for isospectrality imply that the spin-orbit and Darwin terms of
either the upper component or the lower component of the Dirac
spinor vanish, making it equivalent, as far as energy is concerned,
to a spin-0 state. In this case, besides energy, a scalar particle
will also have the same orbital angular momentum as the (conserved)
orbital angular momentum of either the upper or lower component of
the corresponding spin-1/2 particle. We point out a few possible
applications of this result.
\end{abstract}
\maketitle



When describing some strong interacting systems it is often useful,
because of simplicity, to approximate the behavior of relativistic
spin-1/2 particles by scalar spin-0 particles obeying the
Klein-Gordon equation. An example is the case of relativistic quark
models used for studying quark-hadron duality because of the added
complexity of structure functions of Dirac particles as compared to
scalar ones. It turns out that some results (e.g., the onset of
scaling in some structure functions) almost do not depend on the
spin structure of the particle \cite{jeschonnek}. In this work we
will give another example of an observable, the energy, whose value
may not depend on the spinor structure of the particle, i.e.,
whether one has a spin-1/2 or a spin-0 particle. We will show that
when a Dirac particle is subjected to scalar and vector potentials
of equal magnitude, it will have exactly the same energy spectrum as
a scalar particle of the same mass under the same potentials. As we
will see, this happens because the spin-orbit and Darwin terms in
the second-order equation for either the upper or lower spinor
component vanish when the scalar and vector potentials have equal
magnitude. It is not uncommon to find physical systems in which
strong interacting relativistic particles are subject to Lorentz
scalar potentials (or position-dependent effective masses) that are
of the same order of magnitude of potentials which couple to the
energy (time components of Lorentz four-vectors). For instance, the
scalar and vector (hereafter meaning time-component of a four-vector
potential) nuclear mean-field potentials have opposite signs but
similar magnitudes, whereas relativistic models of mesons with a
heavy and a light quark, like D- or B-mesons, explain the observed
small spin-orbit splitting by having vector and scalar potentials
with the same sign and similar strengths \cite{gino_mesons}.

It is well-known that all the components of the free Dirac spinor,
\textit{i.e.}, the solution of the free Dirac equation, satisfy the
free Klein-Gordon equation. Indeed, from the free Dirac equation
\begin{equation}
\label{free_Dirac} (i\hbar\gamma^\mu\partial_\mu-mc)\Psi=0
\end{equation}
one gets
\begin{equation}
\label{free_KG_Dirac}
(-i\hbar\gamma^\nu\partial_\nu-mc)(i\hbar\gamma^\mu\partial_\mu-mc)\Psi=
(\hbar^2\partial^\mu\partial_\mu+m^2c^2)\Psi=0 \ ,
\end{equation}
where use has been made of the relation
$\gamma^\mu\gamma^\nu\partial_\mu\partial_\nu=\partial_\mu\partial^\mu$.
In a similar way, for the time-independent free Dirac equation we
would have
\begin{equation}
\label{free_t_ind_Dirac} (c\,\bmath\alpha\cdot\bmath p +\beta mc^2
)\psi=(-i\hbar c\,\bmath\alpha\cdot\nabla+\beta mc^2)\psi=E\psi \ ,
\end{equation}
where, as usual,
$\psi(\bmath r)=\Psi(\bmath r,t)\,\exp{(i\,E\,t/\hbar)}$,
$\bmath\alpha=\gamma^0\bmath\gamma$ and $\beta=\gamma^0$. Then,
by left multiplying Eq.~(\ref{free_t_ind_Dirac}) by
$c\bmath\alpha\cdot\bmath p+\beta mc^2$, one gets the
time-independent free Klein-Gordon equation
\begin{equation}
\label{free_t_ind_KG}
(c^2\bmath{p}^2+m^2c^4)\psi=(-\hbar^2c^2\nabla^2+m^2c^2)\psi=E^2\psi\
,
\end{equation}
where the relation $\{\beta,\bmath\alpha\}=0$ was used. This all
means that the free four-component Dirac spinor, and of course all
of its components, satisfy the Klein-Gordon equation. This is not
surprising, because, after all, both free spin-1/2 and spin-0
particles obey the same relativistic dispersion relation,
$E^2={\bmath p}^2c^2+m^2c^4$, in spite of having different spinor
structures and thus different wave functions. Since there is no
spin-dependent interaction, one expects both to have the same energy
spectrum.


We consider now the case of a spin-1/2 particle subject to a Lorentz
scalar potential $V_s$ plus a vector potential $V_v$. The
time-independent Dirac equation is given by
\begin{equation}
\label{dirac_V_S} [c\,\bmath\alpha\cdot\bmath p
+\beta(mc^2+V_s)]\psi=(E-V_v)\psi
\end{equation}
It is convenient to define the four-spinors
$\psi_\pm=P_\pm\psi=[(I\pm\beta)/2]\,\psi$ such that
\begin{equation}
\psi_+=\left(\begin{array}{c}
\phi\\[2mm] 0
\end{array}\right)\qquad \psi_-=\left(\begin{array}{c}
0\\[2mm] \chi
\end{array}\right) \ ,
\end{equation}
where $\phi$ and $\chi$ are respectively the upper and lower
two-component spinors. Using the properties and anti-commutation
relations of the matrices $\beta$ and $\bmath\alpha$ we can apply
the projectors $P_\pm$ to the Dirac equation (\ref{dirac_V_S}) and
decompose it into two coupled equations for $\psi_+$ and $\psi_-$:
\begin{eqnarray}
\label{psi-1}
  c\,\bmath\alpha\cdot\bmath p\,\psi_-+(mc^2+V_s)\psi_+ &=&(E-V_v)\psi_+ \\
\label{psi+1}
 c\,\bmath\alpha\cdot\bmath
 p\,\psi_+-(mc^2+V_s)\psi_-&=&(E-V_v)\psi_- \ .
\end{eqnarray}
Applying the operator $c\,\bmath\alpha\cdot\bmath p$ on the left of
these equations and using them to write $\psi_+$ and $\psi_-$ in
terms of $\bmath\alpha\cdot\bmath p\,\psi_-$ and
$\bmath\alpha\cdot\bmath p\,\psi_+$ respectively, we finally get
second-order equations for $\psi_+$ and $\psi_-$:
\begin{eqnarray}
 c^2\bmath p^2\,\psi_++c^2\,\frac{[\bmath\alpha\cdot\bmath p\,\Delta]
 \bmath\alpha\cdot\bmath p\, \psi_+}{E-\Delta+mc^2}
  &=&(E-\Delta+mc^2)(E-\Sigma-mc^2)\psi_+ \\
 c^2\bmath p^2 \,\psi_-+c^2\,\frac{[\bmath\alpha\cdot\bmath p\,\Sigma]
 \bmath\alpha\cdot\bmath p\, \psi_-}{E-\Sigma-mc^2}
 &=&(E-\Delta+mc^2)(E-\Sigma-mc^2)\psi_-
\end{eqnarray}
where the square brackets $[\ ]$ mean that the operator
$\bmath\alpha\cdot\bmath p$ only acts on the potential in front of
it and we defined $\Sigma=V_v+V_s$ and $\Delta=V_v-V_s$. The second
term in these equations can be further elaborated noting that the
Dirac $\alpha_i$ matrices satisfy the relation $
\alpha_i\alpha_j=\delta_{ij}+\frac2\hbar\,i\epsilon_{ijk} S_k$ where
$S_k$, $k=1,2,3$, are the spin operator components.
The second-order equations read now
\begin{eqnarray}
\label{2nd_psi_+}
 c^2 \,\bmath p^2\,\psi_++c^2\,\frac{[\bmath p\,\Delta]
\cdot\bmath p\, \psi_+ + \frac{2i}{\hbar}\,[\bmath
p\,\Delta]\times\bmath p\cdot \bmath S\, \psi_+}{E-\Delta+mc^2}\,
  &=&(E-\Delta+mc^2)(E-\Sigma-mc^2)\psi_+  \\
  \label{2nd_psi_-}
 c^2\,\bmath p^2 \,\psi_-+c^2\,\frac{[\bmath p\,\Sigma]
\cdot\bmath p\, \psi_- + \frac{2i}{\hbar}\,[\bmath
p\,\Sigma]\times\bmath p\cdot \bmath S\, \psi_-}{E-\Sigma-mc^2}\,
 &=&(E-\Delta+mc^2)(E-\Sigma-mc^2)\psi_- .
\end{eqnarray}

Now, if $\bmath p\,\Delta=0$, meaning that $\Delta$ is constant or
zero (if $\Delta$ goes to zero at infinity, the two conditions are
equivalent), then the second term in eq. (\ref{2nd_psi_+})
disappears and we have
\begin{equation}
c^2\,\bmath p^2 \psi_+
  =(E-\Delta+mc^2)(E-\Sigma-mc^2)\psi_+=[(E-V_v)^2-(mc^2+V_s)^2]\psi_+\  ,
\end{equation}
which is precisely the time-independent Klein-Gordon equation for a
scalar potential $V_s$ plus a vector potential $V_v$\footnote{There
are some authors who introduce a scalar potential ${\cal V}_s$ in
the Klein-Gordon equation by making the replacement $m^2c^4\to
m^2c^4+{\cal V}_s^2$. Here we introduce it, as most authors do, as
an effective mass $m^{*\,2}=(m+V_s/c^2)^2$, since it is the way that
it is introduced in the Dirac equation. The two potentials are
related by ${\cal V}_s^2=(mc^2+V_s)^2-m^2c^4$.}. Since the
second-order equation determines the eigenvalues for the spin-1/2
particle, this means that when $\bmath p\,\Delta=0$, a spin-1/2 and
a spin-0 particle with the same mass and subject to the same
potentials $V_s$ and $V_v$ will have the same energy spectrum,
including \textit{both} bound and scattering states. This last
sufficient condition for isospectrality can be relaxed to demand
that just the combination $mc^2+V_s$ be the same for both particles,
\textit{allowing them to have different masses}. This is so because
this weaker condition does not change the gradient of $\Delta$ and
$\Sigma$ and therefore the condition $\bmath p\,\Delta=0$ will still
hold. On the other hand, if the scalar and vector potentials are
such that $\bmath p\,\Sigma=0$, we would obtain a Klein-Gordon
equation for $\psi_-$, and again the spectrum for spin-0 and
spin-1/2 particles would be the same, provided they are subjected to
the same vector potential and $mc^2+V_s$ is the same for both
particles. If both $V_s$ and $V_v$ are central potentials,
\textit{i.e.}, only depend on the radial coordinate, then the
numerators of the second terms in equations (\ref{2nd_psi_+}) and
(\ref{2nd_psi_-}) read
\begin{eqnarray}
 [\bmath p\,\Delta] \cdot\bmath p\, \psi_+ +
\frac{2i}{\hbar}\,[\bmath p\,\Delta]\times\bmath p\cdot \bmath S\,
\psi_+&=&-\hbar^2\Delta'\,\frac{\partial\psi_+}{\partial r}+\frac
2r\,\Delta'\bmath L\cdot \bmath S\,\psi_+  \\
\lbrack\bmath p\, \Sigma\rbrack \cdot\bmath p\, \psi_- +
\frac{2i}{\hbar}\,\lbrack\bmath p\,\Sigma\rbrack\times\bmath p\cdot
\bmath S\, \psi_-&=&-\hbar^2\Sigma'\,\frac{\partial\psi_-}{\partial
r}+\frac 2r\,\Sigma'\bmath L\cdot \bmath S\,\psi_-\ ,
\end{eqnarray}
where $\Delta'$ and $\Sigma'$ are the derivatives with respect to
$r$ of the radial potentials $\Delta(r)$ and $\Sigma(r)$, and
$\bmath L=\bmath r\times\bmath p$ is the orbital angular momentum
operator. From these equations ones sees that these terms, which set
apart the Dirac second-order equations for the upper and lower
components of the Dirac spinor from the Klein-Gordon equation and
thus are the origin of the different spectra for spin-1/2 and spin-0
particles, are composed of a derivative term, related to the Darwin
term which appears in the Foldy-Wouthuysen expansion, and a $\bmath
L\cdot \bmath S$ spin-orbit term. If $\Delta'=0$ ($\Sigma'=0$), then
there is no spin-orbit term for the upper (lower) component of the
Dirac spinor. In turn, since the second-order equation determines
the energy eigenvalues, this means that the orbital angular momentum
of the respective component is a good quantum number of the Dirac
spinor.
This can be a bit surprising, since one knows that in general the
orbital quantum number is not a good quantum number for a Dirac
particle, since $\bmath L^2$ does not commute with a Dirac
Hamiltonian with radial potentials. The reason why this does not
happen in these cases was reported in Refs.~\cite{levi,gin_rep}, and
we now review it in a slight different fashion. Let us consider in
more detail the case of spherical potentials such that $\Delta'=0$.
One knows that a spinor that is a solution of a Dirac equation with
spherically symmetric potentials can be generally written as
\begin{equation}
\label{psi_jm}
 \psi_{jm}(\mbox{\boldmath $r$})=\left(
\begin{array}{c}
\displaystyle {\rm i}\frac{g_{j\,l}(r)}{r} \mathcal{Y}_{j\,l\,m}(%
\mbox{\boldmath $\hat{r}$}) \\[.2cm]
\displaystyle\frac{f_{j\,\tilde l}(r)}{r} \mathcal{Y}_{j\,\tilde l\,m}(%
\mbox{\boldmath $\hat{r}$})
\end{array}
\right)\, .
\end{equation}
where $\mathcal{Y}_{j\,l\,m}$ are the spinor spherical harmonics.
These result from the coupling of spherical harmonics and
two-dimensional Pauli spinors $\chi_{m_s}$,
$\mathcal{Y}_{j\,l\,m}=\sum_{m_s}\sum_{m_l}\clebschgordan{l}{m_l}{1/2}{m_s}{j}{m}
Y_{l\,m_l}\chi_{m_s}$,
where $\clebschgordan{l}{m_l}{1/2}{m_s}{j}{m}$ is a Clebsch-Gordan
coefficient and $\tilde l=l\pm1$, the plus and minus signs being
related to whether one has aligned or anti-aligned spin,
\textit{i.e.}, $j=l\pm1/2$. The spinor spherical harmonics for the
lower component satisfy the relation
$\mathcal{Y}_{j\,\tilde l\,m}=-\bmath\sigma\cdot\bmath{\hat r}\,
\mathcal{Y}_{j\,l\,m}$.
The fact that the upper and lower components have different orbital
angular momenta is related to the fact, mentioned before, that $\bmath L^2$ does not
commute with the Dirac Hamiltonian
\begin{equation}
\label{H_P+_P-} H=c\,\bmath\alpha\cdot\bmath
p+\beta(V_s+mc^2)+V_v=c\,\bmath\alpha\cdot\bmath p+\beta mc^2+\Sigma
P_++\Delta P_-\ ,
\end{equation}
where $P_\pm$ are the projectors defined above. However, when
$\Delta'=0$, there is an extra SU(2) symmetry of $H$ (so-called
``spin symmetry") as first shown by Bell and Ruegg \cite{bell}. When
we have spherical potentials, Ginocchio showed that there is an
additional SU(2) symmetry (for a recent review see \cite{gin_rep}).
The generators of this last symmetry are
\begin{equation}
\label{cal_L}
 \bmath{\mathcal{L}}=\bmath L
P_++\frac1{p^2}\bmath\alpha\cdot\bmath p\,\bmath
L\,\bmath\alpha\cdot\bmath p\,P_-=
\left(\begin{array}{cc} \bmath L&0\\
0&U_p\,\bmath L\,U_p\end{array}\right)\ ,
\end{equation}
where $U_p=\bmath\sigma\cdot\bmath p/(\sqrt{p^2})$ is the helicity
operator. One can check that $\bmath{\mathcal{L}}$ commutes with the
Dirac Hamiltonian,
\begin{eqnarray}
\nonumber[H,\bmath{\mathcal{L}}]&=& [c\,\bmath\alpha\cdot\bmath
p,\bmath L P_++\frac1{p^2}\bmath\alpha\cdot\bmath p\,\bmath
L\,\bmath\alpha\cdot\bmath
p\,P_-]+[\Delta,\frac1{p^2}\bmath\alpha\cdot\bmath p\,\bmath
L\,\bmath\alpha\cdot\bmath p]+[\Sigma,\bmath L] \\
&=&[\Delta,\frac1{p^2}\bmath\alpha\cdot\bmath p\,\bmath
L\,\bmath\alpha\cdot\bmath p\, ]=0 \ ,
\end{eqnarray}
where the last equality comes from the fact that $\Delta'=0$. The
Casimir $\bmath{\mathcal{L}}^2$ operator is given by
$\ds\bmath{\mathcal{L}}^2=\bmath L^2
P_++\frac1{p^2}\bmath\alpha\cdot\bmath p\,\bmath
L^2\,\bmath\alpha\cdot\bmath p\,P_-$.
Applying this operator to the spinor $\psi_{jm}$ (\ref{psi_jm}), we
get
\begin{eqnarray}
\nonumber\bmath{\mathcal{L}}^2\psi_{jm}&=&\bmath L^2\psi_{jm}^+
+\frac1{p^2}\bmath\alpha\cdot\bmath p\,\bmath
L^2\,\bmath\alpha\cdot\bmath p\,\psi_{jm}^-
=\hbar^2l(l+1)\psi_{jm}^++ \frac{\bmath\alpha\cdot\bmath p\,c{\bmath
L}^2\,\psi_{jm}^+}{E-\Delta+mc^2}\\
\label{eigen_L^2}
&=&\hbar^2l(l+1)\psi_{jm}^++\hbar^2l(l+1)\psi_{jm}^-=\hbar^2l(l+1)\psi_{jm}
\ ,
\end{eqnarray}
where $\psi_{jm}^{\pm}=P_{\pm}\psi_{jm}$ and we used the relation,
valid when $\Delta'=0$,
$\ds\psi_{jm}^+=(E-\Delta+mc^2)\frac{\bmath\alpha\cdot\bmath p}{c
p^2}\psi_{jm}^-$.
From (\ref{eigen_L^2}) we see that $\psi_{jm}$ is indeed an
eigenstate of $\bmath{\mathcal{L}}^2$. Thus the orbital quantum
number of the upper component $l$ is a good quantum number of the
system when the spherical potentials $V_s(r)$ and $V_v(r)$ are such
that $V_v(r)=V_s(r)+C_\Delta$, where $C_\Delta$ is an arbitrary
constant. Also, according to we have said before, there is a state
of a spin-0 particle subjected to these same spherical potentials
(or, at least, with a scalar potential such that the sum $V_s+mc^2$
is the same) that has the same energy and the same orbital angular
momentum as $\psi_{jm}$. In addition, the wave function of this
scalar particle would be proportional to the spatial part of the
wave function of the upper component.

Note that the generator of the ``spin symmetry''
$\bmath{\mathcal{S}}$ is given by a similar expression as
(\ref{cal_L}) just replacing $\bmath L$ by $\hbar/2\,\bmath\sigma$
\cite{bell,gin_rep}, meaning that $\bmath{\mathcal{S}}^2\equiv
\bmath{S}^2=3/4\,\hbar^2 I$ so that spin is also a good quantum
number, as would be expected. Actually, one can show that the total
angular momentum operator $\bmath J$ can be written as
$\bmath{\mathcal{L}}+\bmath{\mathcal{S}}$, so that $l$, $m_l$
(eigenvalue of $\mathcal{L}_z$), $s=1/2$, $m_s$ (eigenvalue of
$\mathcal{S}_z$) are good quantum numbers. Then, of course, $j$ and
$m=m_l+m_s$ are also good quantum numbers, but only in a trivial
way, because there is no longer spin-orbit coupling. Therefore, in
the spinor (\ref{psi_jm}) one could just replace the spinor
spherical harmonic $\mathcal{Y}_{j\,l\,m}$ by $Y_{l\,m_l}\chi_{m_s}$
and $\mathcal{Y}_{j\,\tilde l\,m}$ by $-\bmath\sigma\cdot\bmath{\hat
r}\, Y_{l\,m_l}\chi_{m_s}$. Note that if $\Delta$ is a
nonrelativistic potential, $\Delta\ll mc^2$ and $\Delta'\ll
m^2c^4/(\hbar c)$, \textit{i.e.}, it is slowly varying over a
Compton wavelength. In this case, the spin-orbit term will also get
suppressed. In fact, the derivative of the $\Delta$ potential is the
origin of the well-known relativistic spin-orbit effect which
appears as a relativistic correction term in atomic physics or in
the $v/c$ Foldy-Wouthuysen expansion (only the derivative of $V_v$
appears because usually no Lorentz scalar potential $V_s$ is
considered, and therefore $\Delta=V_v$).

When $\Sigma'=0$, or $V_v(r)=-V_s(r)+C_\Sigma$, with $C_\Sigma$ an
arbitrary constant, there is again a $SU(2)$ symmetry, usually
called pseudospin symmetry (\cite{bell,gino}) which is relevant for
describing the single-particle level structure of several nuclei.
This symmetry has a dynamical character and cannot be fully realized
in nuclei because in Relativistic Mean-field Theories the $\Sigma$
potential is the only binding potential for nucleons
\cite{alb1,alb2}. For harmonic oscillator potentials this is no
longer the case, since $\Delta$, acting as an effective mass going
to infinity, can bind Dirac particles \cite{gino_oh,nosso}, even
when $\Sigma=0$. As before, in the special case of spherical
potentials, there is another SU(2) symmetry whose generators are
\begin{equation}
\bmath{\tilde \mathcal{L}}=\frac1{p^2}\bmath\alpha\cdot\bmath
p\,\bmath L\,\bmath\alpha\cdot\bmath p\,P_++\bmath L P_-=
\left(\begin{array}{cc} U_p\,\bmath L\,U_p&0\\
0&\bmath L\end{array}\right)\ .
\end{equation}
In the same way as before, applying $\bmath{\tilde \mathcal{L}}^2$
to $\psi_{jm}$, we would find that
$\bmath{\tilde \mathcal{L}}^2\psi_{jm}=\hbar^2\tilde l(\tilde
l+1)\psi_{jm}$,
that is, this time it is the orbital quantum number of the lower
component $\tilde l$ which is a good quantum number of the system
and can be used to classify energy levels. Again, provided the
vector and scalar potentials are adequately related, there would be
a corresponding state of a spin-0 particle with the same energy and
same orbital angular momentum $\tilde l$, and, furthermore, its wave
function would be proportional to the spatial part of the wave
function of the lower component. As before, the pseudospin symmetry
generator $\bmath{\tilde \mathcal{S}}$ can be obtained from
$\bmath{\tilde \mathcal{L}}$ by replacing $\bmath L$ by
$\hbar/2\,\bmath\sigma$. The good quantum numbers of the system
would be, besides $\tilde l$, $m_{\tilde l}$, $\tilde s\equiv s=1/2$
and $m_{\tilde s}$. Again, $\bmath
J=\bmath{\tilde\mathcal{L}}+\bmath{\tilde\mathcal{S}}$. It is
interesting that, as has been noted by Ginocchio \cite{gino_oh}, the
generators of spin and pseudospin symmetries are related through a
$\gamma^5$ transformation since
$\bmath{\tilde\mathcal{S}}=\gamma^5\bmath{\mathcal{S}}\gamma^5$ and
$\bmath{\tilde\mathcal{L}}=\gamma^5\bmath{\mathcal{L}}\gamma^5$.
This property was used in a recent work to relate spin symmetric and
pseudospin symmetric spectra of harmonic oscillator potentials
\cite{Castro_OH1+1}. There it was shown that for massless particles
(or ultrarelativistic particles) the spin- and pseudo-spin spectra
of Dirac particles are the same. In addition, this means that
spin-symmetric massless eigenstates of $\gamma^5$ would be also
pseudo-spin symmetric and vice-versa. Since in this case
$\Delta=\Sigma=0$, or $V_v=V_s=0$, this is, of course, just another
way of stating the well-known fact that free massless Dirac
particles have good chirality.

Naturally, for free spin-1/2 particles described by spherical waves,
\textit{both} $l$ and $\tilde l$ are good quantum numbers, which
just reflects the fact that one can have free spherical waves with
any orbital angular momentum for the upper or lower component and
still have the same energy, as long as their linear momentum
magnitude is the same, or, put in another way, the energy of a free
spin-1/2 particle cannot depend on its direction of motion.


In summary, we showed that when a relativistic spin-1/2 particle is
subject to vector and scalar potentials such that $V_v=\pm
V_s+C_\pm$, where $C_\pm$ are constants, its energy spectrum does
not depend on their spinorial structure, being identical to the
spectrum of a spin-0 particle which has no spinorial structure. This
amounts to say that if the potentials have these configurations
there is no spin-orbit coupling and Darwin term. If the scalar and
vector potentials are spherical, one can classify the energy levels
according to the orbital angular momentum quantum number of either
the upper or the lower component of the Dirac spinor. This would
then correspond to having a spin-0 particle with orbital angular
momentum $l$ or $\tilde l$, respectively.
This spectral identity can of course happen only with potentials
which do not involve the spinorial structure of the Dirac equation
in an intrinsic way. For instance, a tensor potential of the form
${\rm i}\beta\sigma^{\mu\nu}(\partial_\mu A_\nu-\partial_\nu A_\mu)$
does not have an analog in the Klein-Gordon equation, so that one
could not have a spin-0 particle with the same spectrum as a
spin-1/2 particle with such a potential. This is the case of the
so-called Dirac oscillator \cite{dirac_osc} (see \cite{nosso} for a
complete reference list), in which the Dirac equation contains a
potential of the form ${\rm i}\beta\sigma^{0i}m\omega r_i={\rm i}
m\omega\beta\bmath\alpha\cdot\bmath r$. Another important potential,
the electromagnetic vector potential $\bmath A$, which is the
spatial part of the electromagnetic four-vector potential, can be
added via the minimal coupling scheme to both the Dirac and the
Klein-Gordon equations. Since $\bmath\alpha\cdot(\bmath p-e\bmath
A)\bmath\alpha\cdot(\bmath p-e\bmath A)= (\bmath p-e\bmath
A)^2+2e\hbar\nabla\times\bmath A\cdot\bmath S$, the spectra of
spin-0 and spin-1/2 particles cannot be identical as long as there
is a magnetic field present, even though the condition $V_v=\pm
V_s+C_\pm$ is fulfilled. It is important also to remark that, since
for an electromagnetic interaction $V_v$ is the time-component of
the electromagnetic four-vector potential, this last condition is
gauge invariant in the present case, in which we are dealing with
stationary states, \textit{i.e}, time-independent potentials. So, in
the absence of a external magnetic field (allowing, for instance, an
electromagnetic vector potential $\bmath A$ which is constant or a
gradient of a scalar function), a spin-0 and spin-1/2 particle
subject to the same electromagnetic potential $V_v$ and a Lorentz
scalar potential fulfilling the above relation would have the same
spectrum.

The remark made above about the similarity of spin-0 and spin-1/2
wave functions can be relevant for calculations in which the
observables do not depend on the spin structure of the particle,
like some structure functions. One such calculation was made by
Paris \cite{Paris} in a massless confined Dirac particle, in which
$V_v=V_s$. It would be interesting to see how a Klein-Gordon
particle would behave under the same potentials. More generally,
this spectral identity can also have experimental implications in
different fields of physics, since, should such an identity be
found, it would signal the presence of a Lorentz scalar field having
a similar magnitude as that of a time-component of a Lorentz vector
field, or at least differing just by a constant.

\vskip1cm
\begin{acknowledgments}
We acknowledge financial support from CNPQ, FAPESP and FCT (POCTI)
scientific program.
\end{acknowledgments}

\end{document}